\documentstyle [12pt]{article}
\begin{document}
\input epsf

\hfill {WM-00-113}

\hfill {\today}

\vskip 1in   \baselineskip 24pt

{
\Large
   \bigskip
   \centerline{t-channel production of heavy charged leptons}
 }

\vskip .8in
\def\bar{\overline}

\centerline{Shuquan Nie\footnote{Email: sxnie@physics.wm.edu} and Marc
Sher\footnote{Email: sher@physics.wm.edu}  }
\bigskip
\centerline {\it Nuclear and Particle Theory Group}
\centerline {\it Physics Department}
\centerline {\it College of William and Mary, Williamsburg, VA 23187, USA}

\vskip 1in

{\narrower\narrower
 We study the pair production of heavy charged exotic leptons at $e^{+}e^{-}$
colliders  in the $SU(2)_L \times SU(2)_I \times U(1)_Y $ model.  This gauge group is a
subgroup of the grand  unification group $E_6$; $SU(2)_I$ commutes with the electric
charge operator, and the three corresponding gauge  bosons are electrically neutral.  In
addition to  the standard
$\gamma$ and Z boson contributions, we  also include the contributions from extra
neutral gauge bosons.   A t-channel contribution due to $W_I$-boson exchange, which is
unsuppressed by mixing angles, is quite important. We calculate the left-right and
forward-backward asymmetries, and discuss how to differentiate different models. }

\section{Introdution}

Many extensions of the Standard Model (SM) contain exotic fermions.  Strongly
interacting exotics, such as heavy quarks, can be produced in abundance at the Tevatron
or the LHC.  However, particles which are not strongly interacting, such as heavy
charged leptons, can best be produced at an electron-positron collider.   In general,
studies of heavy charged leptons at such colliders focus on s-channel production,
through a $\gamma$, $Z$, $Z'$, etc.   The phenomenology of exotic particles has been
considered widely [1-8]. A good report can be found in  Ref.
\cite{Hewett1}.

In this paper, we note that a model which arises from superstring-inspired $E_6$ grand
unification models will allow pair production of heavy charged leptons in the t-channel.  We
discuss this model, and study the forward-backward and left-right asymmetries at linear
colliders. For simplicity, we neglect mixing between extra
particles (bosons or fermions) and the  normal particles of the SM, since such mixing
angles are generally small.

\section{The model} There are many phenomenologically acceptable low energy models
which arise from $E_6$.
\begin{eqnarray} (a) E_6 &\longrightarrow& SU(3)_C \times SU(2)_L \times U(1)_Y \times
U(1)_{Y^{\prime}} \nonumber \\ (b) E_6 &\longrightarrow& SO(10) \times U(1)_{\psi}
\longrightarrow SU(5) \times U(1)_{\chi}\times U(1)_{\psi} \nonumber \\ (c) E_6 &\longrightarrow& SU(3)_C
\times SU(2)_L \times SU(2)_R \times U(1)_L \times U(1)_R \nonumber \\ (c^{\prime}) E_6
&\longrightarrow& SU(3)_C \times SU(2)_L \times SU(2)_I \times U(1)_Y \times
U(1)_{Y^{\prime}} \nonumber
\end{eqnarray} where $U(1)_{\psi}$ and $U(1)_{\chi}$ can be combined into
$U(1)_{\theta}$ in model (b), reducing it to the effective  rank-5 model
$SU(3)_C \times SU(2)_L \times U(1)_Y \times U(1)_{\theta}$, which is  most often considered
in the literature. Models (c) and $(c^{\prime})$ come from the subgroup $SU(3)_C \times
SU(3)_L \times SU(3)_R$.  The {\bf 27}-dimensional fundamental representation has the
branching rule
\begin{equation} {\bf 27}=\underbrace{({\bf 3}^c,{\bf 3},{\bf 1})}_{q}+\underbrace{
(\bar{{\bf 3}}^c, {\bf 1}, \bar{{\bf 3}})}_{\bar{q}} +\underbrace{({\bf 1}^c, \bar{{\bf
3}}, {\bf 3})}_{l}
\end{equation} and the particles of the first family are assigned as
\[
  \left( \begin{array}{c} u\\d\\h \end{array} \right)+
  \left( \begin{array}{lcr} u^c & d^c & h^c \end{array} \right)+
  \left( \begin{array}{ccc} E^c & \nu & N \\ N^c & e & E\\e^c & \nu^c & S^c
  \end{array} \right)  \] where $SU(3)_L$ operates vertically and $SU(3)_R$ operates
horizontally. (Different symbols for these particles may be used in the literature.)

The most common method of breaking the $SU(3)_R$ factor is to break the {\bf 3} of
$SU(3)_R$ into {\bf 2}+{\bf 1}, so that $(u^c, d^c)$ forms an $SU(2)_R$ doublet with
$h^c$ as a $SU(2)_R$ singlet. This gives Model (c), the familiar left-right  symmetric
model \cite{Candelas}. Model (c) can be reduced further to an effective rank-5  model
with $U(1)_{V=L+R}$. Another possibility, resulting in Model $(c^{\prime})$, is to
break the {\bf 3} of the $SU(3)_R$ into ${\bf 1} + {\bf 2}$ so  that $(d^c, h^c)$ forms
an $SU(2)$ doublet with  $u^c$ as a singlet. In this option, the $SU(2)$ doesn't
contribute to  the electromagnetic charge operator and it is called $SU(2)_I$ (I stands
for Inert).   Then the vector gauge bosons corresponding to $SU(2)_I$ are neutral.
Model $(c^{\prime})$ can be reduced to an effective rank-5 model $SU(3)_C \times
SU(2)_L \times U(1)_Y \times SU(2)_I$. Both of them will be considered in this paper.

At the  $SU(2)_L \times SU(2)_I \times U(1)_Y
\times U(1)_{Y^{\prime}}$ level, a single generation of fermions can be represented as
\[ \left( \begin{array}{cc} \nu & N\\ e^{-} & E^{-}  \end{array} \right)_L, \hspace{2mm}
   \left( \begin{array}{c} u \\ d \end{array} \right)_L, \hspace{2mm}
   \left( \begin{array}{lr} d^c & h^c \end{array} \right)_L, \hspace{2mm}
   \left( \begin{array}{c} E^c \\ N^c \end{array} \right)_L, \hspace{2mm}
   \left( \begin{array}{lr} \nu^c & S^c \end{array} \right)_L, \hspace{2mm}
   h_L, \hspace{2mm} e_L^c, \hspace{2mm} u^c_L
\] where $SU(2)_{L(I)}$ acts vertically (horizontally).  Note that additional
heavy leptons $\left({N\atop E}\right)$ and its conjugate $\left({E^c\atop N^c}\right)$
form two new isodoublets under
$SU(2)_L$.

\section{Cross Section Production and Asymmetries} The relevant interactions for the
process $e^+e^- \longrightarrow E^+E^-$ are
\begin{eqnarray} {\cal L} &=& \sum _{f=e,E} Q_f {\bar f}_{\alpha} \gamma ^{\mu} f_{\alpha} A_{\mu}
 +\frac{g}{\cos{\theta_W}} \bar{e}_{\alpha} \gamma^{\mu}
(T^3_{e_{\alpha}}-Q_e \sin^2{\theta_W}) e_{\alpha} Z_{\mu} \nonumber \\
            & &  +\frac{g}{2 \cos{\theta_W}} \bar{E}_{\alpha} \gamma^{\mu} (1-2
\sin^2{\theta_W}) E_{\alpha} Z_{\mu}  \nonumber \\
         & & +\frac{g_I}{2 \sqrt{2}} \bar{e} \gamma^{\mu} (1- \gamma_5) E W_{I \mu}
+H.c. \nonumber \\
         & & +\frac{g_I}{4} (\bar{E} \gamma^{\mu} (1-\gamma_5) E -\bar{e} \gamma^{\mu}
(1-\gamma_5) e) Z_{I \mu} \nonumber \\
         & & + \sum_{f=e,E} g_{Y^{\prime}} \frac{Y^{\prime}_{f_{\alpha}}}{2}
\bar{f}_{\alpha} \gamma^{\mu} f_{\alpha} Z^{\prime}_{\mu}
\end{eqnarray}      where $\alpha = L$ or $R$. $g$, $g_I$ and $g_{Y^{\prime}}$ are coupling constants 
and $\theta_W$ is the electroweak mixing angle. For simplicity, we will assume that
$g_I=g$ and $g_{Y^{\prime}} =g_Y$ in our numerical results, it is straightforward to
relax this assumption. The first two lines are couplings between fermions and standard
$\gamma$ and Z. The rest are couplings with extra neutral gauge bosons. The
$e^+e^-\rightarrow E^+E^-$ process can proceed via s-channel exchange of a $\gamma, Z,
Z'$ or $Z_I$, and can also proceed via t-channel exchange of a $W_I$.  Each amplitude can
be written as the form of
\begin{equation} C_i \bar{v}_e \gamma^{\mu} (1-a_i \gamma_5) u_e \bar{u}_E
\gamma_{\mu} (1-b_i \gamma_5) v_E.
\end{equation}

Note that the $W_I$ leads to a t-channel process unsuppressed by small mixing angles.
This is unique to this model.  Note that  if one considered production of the heavy
charged leptons which form an $SU(2)_I$ doublet with the muon or the tau, then the
processes would be identical except that the t-channel process would be absent.

The differential cross section for this process is given by
\begin{equation}
\frac{d \sigma}{d \cos{\theta}}= \frac{1}{8 \pi s } \sqrt{\frac{1}{4}-\frac{m_E^2 }{s}}
\{ D_1 (m_E^2 -u)^2+D_2 (m_E^2-t)^2 + 2 D_3 m_E^2 s \}
\end{equation} s, t and u are the Mandelstam variables, and with
\begin{eqnarray} D_1 &=& \sum_{i,j=1}^5 C_i C_j \{ (1+a_i a_j) (1+b_i
b_j)+(a_i+a_j)(b_i+b_j) \} \nonumber \\ D_2 &=& \sum_{i,j=1}^5 C_i C_j \{ (1+a_i a_j)
(1+b_i b_j)-(a_i+a_j)(b_i+b_j) \} \nonumber \\ D_3 &=& \sum_{i,j=1}^5 C_i C_j \{ (1+a_i
a_j) (1-b_i b_j) \} 
\end{eqnarray} where the $C_i$, $a_i$ and $b_i$ are given in Table 1.
\begin{center}
Table 1 \hspace{0.4 cm}  Coefficients appearing in Eq. (5) \\ \vspace{0.3 cm}
\begin{tabular}{|c|c|c|c|}
\hline \hline
i & $C_i$ & $a_i$ & $b_i$ \\ \hline
1 &$\frac{e^2}{s}$&0	   &0\\	   \hline
2 &$\frac{g^2(1-4\sin^2 \theta_W)(1-2 \sin^2 \theta_W)}{8 \cos^2 \theta_W(s-m^2_Z)}$ &$\frac{1}{1-4
 \sin^2 \theta_W}$&0 \\ \hline
3 &$\frac{-g_I^2}{16 (s-m^2_{Z_I})}$&1&1 \\ \hline
4 &$\frac{-9g^2_{Y^{\prime}}}{144 (s-m^2_{Z^{\prime}})}$&$-\frac{1}{3}$&$-\frac{1}{5}$ \\ \hline
5 &$\frac{g^2_I}{8 (t-m^2_{W_I})}$ &1&1 \\
\hline \hline
\end{tabular}
\end{center}

The forward-back asymmetry is defined by
\begin{equation} A_{FB} =\frac{\int_0^1 \frac{d\sigma}{d\cos{\theta}}
d\cos{\theta}-\int_{-1}^0 \frac{d\sigma}{d\cos{\theta}} d\cos{\theta}}{\int_{-1}^1
\frac{d\sigma}{d\cos{\theta}} d\cos{\theta}}
\end{equation}
and the left-right asymmetry is defined by
\begin{equation} A_{LR}=\frac{\sigma_L-\sigma_R}{\sigma_L+\sigma_R},
\end{equation}
Note that the $C_i$, $a_i$ and $b_i$ will be somewhat different for $\sigma_L$ and
$\sigma_R$ due to the insertion of the projection operator in Eq. (3).  Both
$A_{FB}$ and
$A_{LR}$ at
$e^+e^-$ colliders were studied in Ref.
\cite{Rizzo, Hewett2}, but only s-channel contributions were considered. 
  
\section{Results}  The electroweak part $SU(2)_L \times U(1)_Y$ has
been measured precisely. Let us first consider the rank 5 case. Setting $g_{Y'}=0$, we have two gauge
boson mass parameters $m_{W_I}$ and $m_{Z_I}$.  We
will assume that these masses are equal and thus there is only one mass
parameter remaining, which we choose to be near the experimental lower
bound for direct production\cite{abe}, $m_{Z_I}=650$ GeV.  This is basically the same as
assuming that the gauge bosons do not substantially mix with each other.  The numerical
results for cross section, forward-backward and left-right asymmetries are shown in
Figs. 1-3.  We have plotted the results for $E^+E^-$ and $M^+M^-$ production, where
$M$ is the
$SU(2)_I$ partner of the muon or tau (the only difference will be due to the t-channel
process). For comparison, we also include the standard model results for both a
vectorlike heavy lepton and a chiral heavy lepton.  Although we have assumed that the
$Z_I$ mass ($E,M$ mass) is $650$ GeV ($200$ GeV), it is easy to see how the figures will
be qualitatively modified if these assumptions are relaxed.

In the rank 6 model, one has an additional mass scale and additional coupling.  If we
assume that the $g_{Y'}$ coupling is the same as $g_Y$, and that the mass of the $Z'$
is ${5 g_Y\over 3 g_I}M_{Z_I}$, then one can recalculate the cross section,
forward-backward and left-right asymmetries.  We find that there is not a
substantial difference from the rank 5 case, except in the immediate vicinity of the $Z'$
mass.

\section{Conclusions}

How does one detect these leptons?  The main decay modes depend sensitively on the masses
and mixing angles.  Since the $E$ and its isodoublet partner $N$ are degenerate in the
limit of no mixing, one expects the $E\rightarrow NW^*$ to be into a virtual $W$, leading
to a three-body decay.  Since the allowed three-body phase space is very small, this
decay will be negligible unless the mixing with the lighter generations is extremely
small.  In the more natural case, in which such mixing is not very small, the two-body
decays
$E\rightarrow
\nu_eW$ and
$E\rightarrow eZ$ would dominate.  A
detailed analysis of the lifetimes and the decay modes can be found in Ref.
\cite{frampton}.  There, it was shown that the ratio of $\Gamma(E\rightarrow
e Z)$ to $\Gamma(E\rightarrow \nu_e W)$ is given by the ratio of $|U_{Ee}|^2$ to
$|U_{E\nu_e}|^2$.   This is very model-dependent.  

 Certainly, the signature for
$E\rightarrow eZ$ would be quite dramatic.  Even if the $Z$ decays hadronically or
invisibly, the monochromatic electron, plus the invariant mass of the $Z$ decay
products, would allow for virtually background-free detection.   The signature for
$E\rightarrow \nu_e W$ is less dramatic, but would lead to $W^+W^-$ plus missing
transverse momentum.  As discussed in Ref. \cite{Montalvo}, requiring that the $W$'s
decay leptonically gives a signal of $l^+l^-$, where $l=(e,\mu)$.  The backgrounds, due
to
$e^+e^-\rightarrow \tau^+\tau^-, W^+W^-$ and $ZZ$, can be eliminated by calculating the
invariant mass of the charged fermion pair.  The signal would be striking since it would
consists of a pair of $l^+l^-$ with approximately the same invariant mass.

Suppose these leptons are found.  One would first learn the cross section.  Unless one is
in the vicinity of the $Z_I$ resonance, the cross section in this model would be somewhat
higher than the standard model.  For example, at an NLC of $\sqrt{s}=500$ GeV and
luminosity of $6\times 10^4$ pb${}^{-1}$/yr and for a heavy lepton of $200$ GeV, one
expects approximately
$2\times 10^4$ SM vectorlike fermion pairs produced per year, whereas one has $3\times
10^4$
$E^+E^-$ pairs and
$5\times 10^4$
$M^+M^-$ pairs (note that the t-channel process destructively interferes).   In the
vicinity of the resonance, of course, the cross section can be much larger.  
As discussed in the previous paragraph, if the main decay is into $\nu W$, then a very
clear signature arises if both $W$'s decay into $e\nu_e$ or $\mu\nu_\mu$.  This will
occur approximately $5\%$ of the time, giving a few thousand  such events per year. 
Necessary cuts on the transverse missing energy will reduce the number of usable events,
but it should still be several hundred per year, with very low background.  If the main
decay is into
$eZ$ or $\mu Z$, then the signature is even more dramatic.

There is no forward-backward asymmetry for the pair production of  SM vectorlike fermions,
while the polarization asymmetry for heavy SM chiral fermions is very small.  Therefore,
combining $A_{FB}$ with $A_{LR}$ would make it very straightforward to distinguish
$E^+E^-$ and $M^+M^-$ pairs from SM fermions.  The  behavior of the asymmetries for each
of these is very different at high $\sqrt{s}$.

An important point is to note that the statistical uncertainty, $\left({1-A^2\over
N}\right)^{1/2}$, is very small for this model.  With the approximate number of
reconstructed events being between several hundred and several thousand, this gives a
statistical uncertainty of between $1$ and $10$ percent.  This will be even smaller in
the vicinity of the resonance.  From the figures, it is clear that this uncertainty is
small enough that the various models can be distinguished, even off-resonance.

We thank JoAnne Hewett for a useful conversation. This work was supported by the National
Science Foundation grant NSF-PHY-9900657
\newpage

\def\prd#1#2#3{{\rm Phys. ~Rev. ~}{\bf D#1}, #3 (19#2)}
\def\plb#1#2#3{{\rm Phys. ~Lett. ~}{\bf B#1}, #3 (19#2) }
\def\npb#1#2#3{{\rm Nucl. ~Phys. ~}{\bf B#1}, #3 (19#2) }
\def\prl#1#2#3{{\rm Phys. ~Rev. ~Lett. ~}{\bf #1}, #3 (19#2) }

\bibliographystyle{unsrt}

\newpage

\begin{figure}
\centerline{ \epsfysize 4in \epsfbox{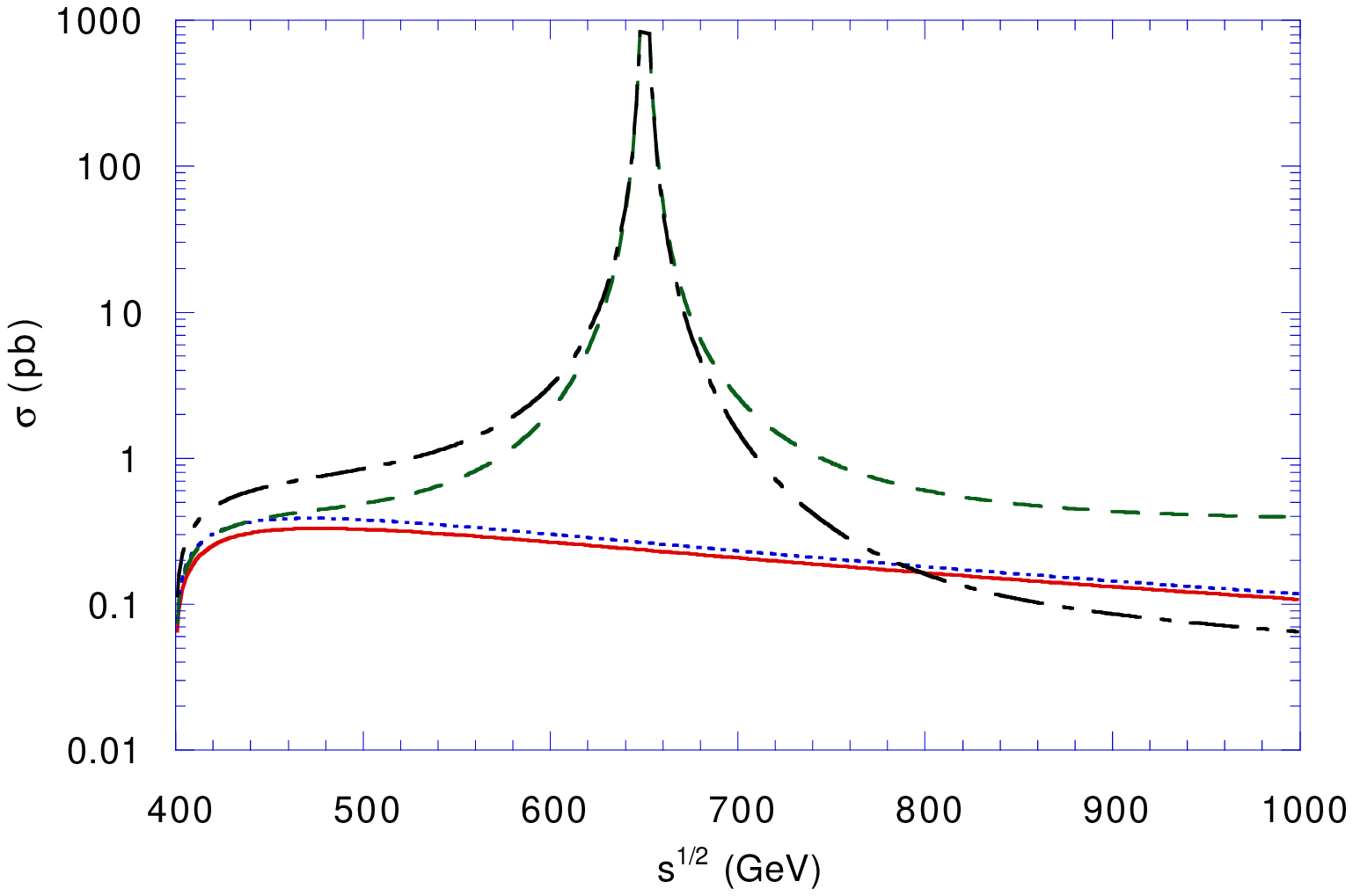}  }
\caption{Total cross section for  the process $e^+e^-\rightarrow L^+L^-$ as a function
of $\sqrt{s}$, for a heavy lepton of $200$ GeV.  The solid and dotted lines correspond
to Standard Model production of chiral and vectorlike fermions, respectively.  The
dashed and dot-dashed lines correspond to $L=E$ and $L=M$ in the $SU(2)_I$ model,
respectively, where
$E$ and $M$ are the $SU(2)_I$ partners of the electron and muon.}
\end{figure}

\begin{figure}
\centerline{ \epsfysize 4in \epsfbox{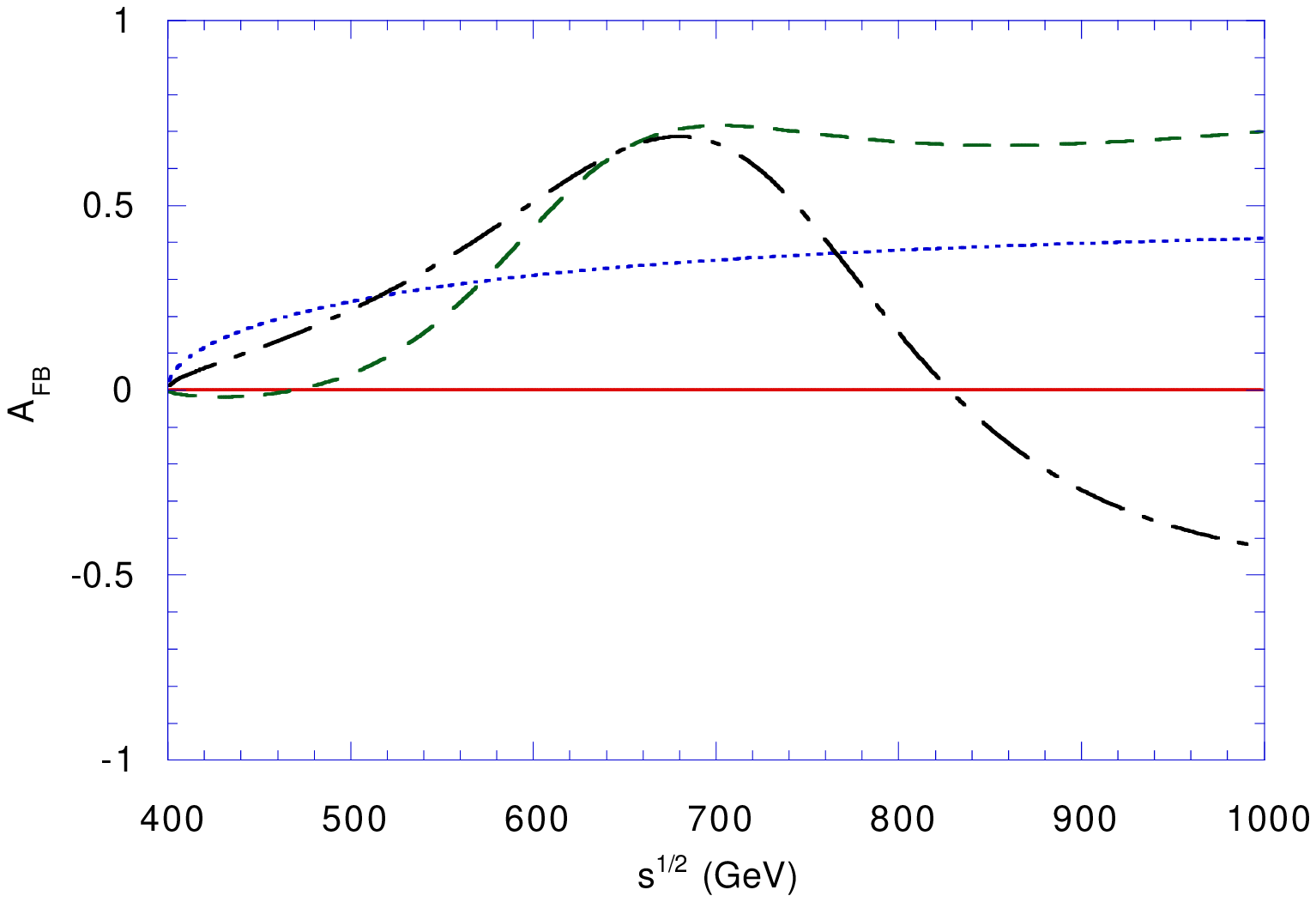}  }
\caption{$A_{FB}$, the forward-backward asymmetry,  for  the
process
$e^+e^-\rightarrow L^+L^-$ as a function of $\sqrt{s}$, for a heavy lepton of $200$
GeV.  The solid and dotted lines correspond to Standard Model production of chiral and
vectorlike fermions, respectively.  The dashed and dot-dashed lines correspond to $L=E$
and $L=M$ in the
$SU(2)_I$ model, respectively, where
$E$ and $M$ are 
the $SU(2)_I$ partners of the electron and muon.}
\end{figure}

\begin{figure}
\centerline{ \epsfysize 4in \epsfbox{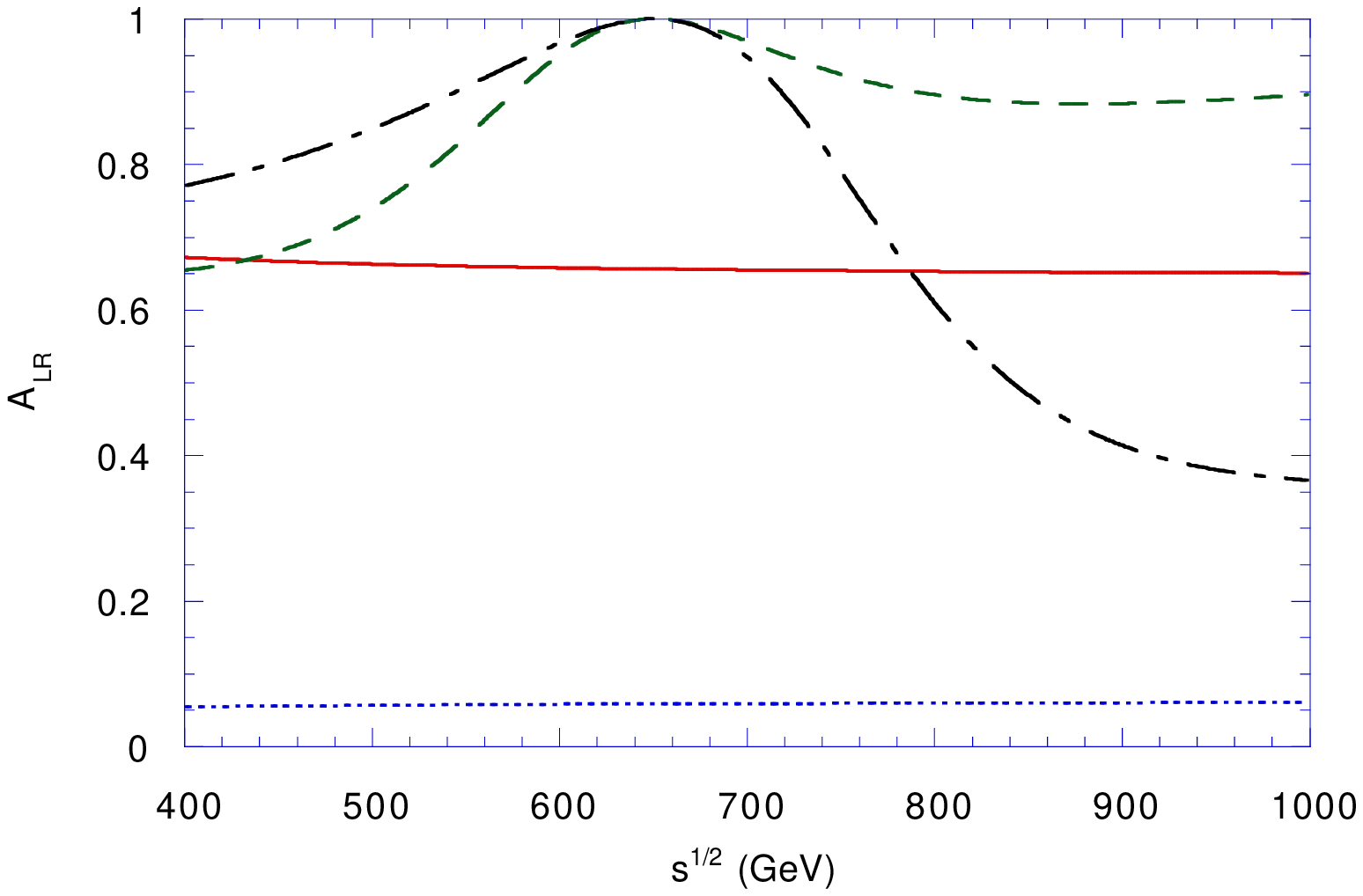}  }
\caption{$A_{LR}$, the left-right asymmetry, for  the process
$e^+e^-\rightarrow L^+L^-$ as a function of $\sqrt{s}$, for a heavy lepton of $200$
GeV.  The solid and dotted lines correspond to Standard Model production of chiral and
vectorlike fermions, respectively.  The dashed and dot-dashed lines correspond to $L=E$
and $L=M$ in the
$SU(2)_I$ model, respectively, where
$E$ and $M$ are the $SU(2)_I$ partners of the electron and muon.}
\end{figure}
\end{document}